\documentstyle[twocolumn,epsf]{jpsj}

\def\runtitle{
Coexistent State of Charge Density Wave and Spin Density Wave in
One-Dimensional Quarter 
Filled Band Systems
}
\def\runauthor{Keita {\sc Kishigi} 
and Yasumasa {\sc Hasegawa} 
}

\title{
Coexistent State of Charge Density Wave and Spin Density Wave in
One-Dimensional Quarter 
Filled Band Systems under Magnetic Fields
}

\author
{
Keita {\sc Kishigi}
\footnote{E-mail: kishigi@sci.himeji-tech.ac.jp}
 and Yasumasa {\sc Hasegawa} 
}

\inst
{
Faculty of Science, Himeji Institute of Technology, Akou-gun, Hyogo
678-1297, Japan 
}

\recdate
{
\today
}

\abst
{
We theoretically study how the 
coexistent state of the charge density wave and 
the spin density wave in the one-dimensional 
quarter filled band is 
enhanced by magnetic fields. 
We found that 
when the correlation between electrons is strong 
the spin density wave state is suppressed under high magnetic fields, 
whereas the charge density wave state still remains. 
This will be observed in experiments such as the X-ray 
measurement. 
}

\kword
{
one-dimensional quarter filled band, charge density wave, spin density wave,
quasi-one-dimensional organic conductors, 
Pauil paramagnetic limit, strongly correlated system 
}

\begin{document}
\sloppy
\maketitle
\section{Introduction}
It is found that 
the system with the one-dimensional quarter filled band 
becomes the coexistent state of the charge 
density wave (CDW) and 
spin density wave (SDW) due to the 
interplay between the on-site Coulomb interaction ($U$) and the 
inter-site Coulomb interaction ($V$) by recent 
theoretical works.\cite{seofukuyama,nobuko,nobuko2,Mazumdar} 
The inter-site Coulomb interaction plays 
important role of the charge ordering, and 
the ground state is the coexistent state of 
$2k_{\rm F}$-SDW and $4k_{\rm F}$-CDW,\cite{seofukuyama} 
where $k_{\rm F}$ is Fermi wave vector, $k_{\rm F}=\pi /4a$ and 
$a$ is the lattice constant. 
Furthermore, when the next nearest neighbor and the dimerization of the 
energy band are considered, it has been indicated that 
$2k_{\rm F}$-SDW and $2k_{\rm F}$-CDW 
coexist.\cite{nobuko,nobuko2} 

Quasi-one dimensional organic conductors 
such as (TMTSF)$_2$$X$ and (TMTTF)$_2$$X$ 
($X$=ClO$_4$, PF$_6$, AsF$_6$, ReO$_4$, Br, SCN, etc.) 
are known as the one-dimensional quarter filled band and 
exhibit many kinds of ground state, 
for example, spin-Peierls, 
SDW, 
superconductivity.\cite{review,jerome} 
In (TMTSF)$_2$PF$_6$, the incommensurate SDW is occurred at $T=12$ K, 
where the wave vector is (0.5, 0.24,-0.06) by 
NMR measurement.\cite{takahashi,delrieu} 
Recently, from the X-ray measurement, Pouget and Ravy argue 
the coexistence of $2k_{\rm F}$-SDW and 
$2k_{\rm F}$-CDW,\cite{pouget} which has been 
theoretically explained by Kobayashi et al.\cite{nobuko,nobuko2} 
mentioned above 
and Mazumdar et al.\cite{Mazumdar}. 

On the other hand, in (TMTTF)$_2$$X$, (X=Br and SCN), 
it is known that the ground state is the antiferromagnetic 
phase understood as Mott-Hubbard inslator phase due to 
the dimerization and the quarter filling. 
It is clear that the wave vector of the SDW 
is commesurate, (0.5, 0.25,0) from the measurements of 
$^{13}$C-NMR\cite{barthel} and $^1$H-NMR\cite{nakamura}. 
From the angle depenence of satelite peak positions of 
$^1$H-NMR\cite{nakamura,nakamura2}, 
the alignment of the spin moment along the conductive axis (a-axis) 
becomes ($\uparrow,0,\downarrow,0$), which 
corresponds to the recent calculational results\cite{seofukuyama,tanemura}. 
In (TMTTF)$_2$Br, $4k_{\rm F}$-CDW accompanied by $2k_{\rm F}$-SDW 
is found in X-ray measurments.\cite{pouget} 
This can be explained by Seo and Fukuyama\cite{seofukuyama} 
by using the extended Hubbard model.

When the pressure is applied, the commensurate antiferromagnetic phase 
in (TMTTF)$_2$Br at 
ambient pressure changes to 
the incommensurate SDW phase such as (TMTSF)$_2$$X$.\cite{klemme}
It is originated to the increasing of the hopping transfer integral, $t$. 
In the case of small $t$, as the exchange interaction is strongly
influenced, the state becomes Mott antiferromagnetic state. 
When $t$ becomes large, the system becomes the SDW phase due to 
the Peierls instability of the Fermi surface. 
The difference between 
(TMTTF)$_2$$X$ and (TMTSF)$_2$$X$ is whether the 
charge or spin ordering is localized or not. 
This difference is attributed that 
$U/t$ in (TMTTF)$_2$$X$ is 
larger than that in (TMTSF)$_2$$X$. 
It is indicated that $U/t\simeq 5.0$ (3.0) in (TMTTF)$_2$$X$ 
((TMTSF)$_2$$X$) since $t$ in (TMTTF)$_2$$X$ ((TMTSF)$_2$$X$) are 
about 0.2 (0.3) eV by the extended Huckel band calculations.
\cite{band,band2,band3,band4}
We consider the system in (TMTTF)$_2$$X$ ((TMTSF)$_2$$X$) 
as strongly (non-strongly) correlated. 

In the one-dimensional system, 
when the magnetic field ($H$) is applied to 
$c$-axis, the amplitudes of the charge density or 
the spin moment along the $c$-axis in the CDW or SDW state by coupling of
electrons with same spins are suppressed due to 
Pauli paramagnetic limit field ($H_p$),\cite{pauli,pauli2,Mckenzie} 
where $H_p\simeq \Delta (0)/\sqrt{2}\mu_{\rm B}$, 
$\Delta(0)$ is the amplitude of the 
energy gap at $H=0$ and 
$\mu_{\rm B}=e\hbar/2m_{0}c$ is the Bohr magneton. 
The energy band is splitted by 
Zeeman effect, so that 
the original wave vector at $H=0$ 
becomes the not good nesting vector. 
In the case of the magnetic field applied along 
the $a$- or $b$-axis, the CDW and SDW are not broken, 
because the nesting vector
is unchanged by Zeeman effect.
In other words, 
the CDW and SDW are not influenced when 
the magnetic field is applied perpendicular to 
the easy axis. 
This picture is for the weak coupling system.

In the strong coupling system, 
the $H$-dependence of the 
antiferromagnetic state by one-dimensional Ising model has been 
studied in the mean field approximation.\cite{ising} 
The amplitude of the spin moment 
along the $c$-axis of the antiferromagnetic state 
under magnetic fields applied along the $a$-axis 
is easily obtained, which 
obey 
\begin{eqnarray}
S_z(H,j)/S_z(0,j)=\sqrt{1-(H/H_x^0)^2},
\end{eqnarray}
where 
$S_z(H,j)$ is the amplitude of the spin moment 
along the $c$-axis at 
$j$ site at $H=0$ and $H_x^0$ is the critical field at which the 
ordering of the antiferromagnetic state disappear.\cite{ising} 
Since the spin of electrons are tilted to 
$a$-axis by the magnetic field, 
$S_z$ is smaller upon increasing $H_x$, 
finally, $S_z$ becomes zero.

In the case of magnetic fields applied to the $c$-axis, 
the antiferromagnetic state along the $c$-axis is kept, 
because if the spin is tilted to $a$-$b$ plane, this tilted state 
is not unstable in weak fields. 
However, the paramagnetic state becomes more
stable 
at higher critical field, $H_z^0$, 
where $H_z^0=H_x^0$, \cite{ising}

We try to analyze 
the coexistent state 
in the two cases when the electron correlation is strong or not, 
since the {\it coexistent} phase of CDW and SDW 
under magnetic fields dose not have been studied 
although the state of CDW or SDW under magnetic fields has been studied. 
When $U/t$ and $V/t$ are large ($\sim 5.0$), 
we consider that the system is strongly correlated, 
because the charge and spin are 
localized as shown in Figs. 1 and 2. 
On the other hand, 
in small $U/t$ and $V/t$ ($\sim 1.5$), the 
correlation between electrons is not strong, 
where there are small amplitudes of charge 
density and spin moment as shown in 
Fig. 7, 8 and 9. 
We calculate to compare the strong coupling system with 
the non-strong coupling system by using of 
two sets of the values of $U/t$ and $V/t$.

In this paper, 
we calculate the self-consistent solutions at $T=0$ for 
the one-dimensional band model under the magnetic field 
perpendicular to (or parallel to) the a-axis 
based on the mean field approximation. 
We use the one-dimensional quarter filled extended 
Hubbard model, where the effect of 
the dimerization do not be considered to 
be simplified problems. 
 
\section{Formulation}

We treat the one-dimensional extended 
Hubbard model, 

\begin{eqnarray}
\hat{\cal H}&=&\hat{\cal K}+\hat{\cal U}+\hat{\cal V}, \\
\hat{\cal K}&=&t\sum_{i,\sigma}(C^{\dagger}_{i,\sigma} C_{i+1,\sigma}+
h.c.)
-\frac{\mu_{\rm B}gH_{j}}{2}\sum_{i, \sigma}n_{i,\sigma},\\
\hat{\cal U}&=&U\sum_{i}n_{i, \uparrow}n_{i, \downarrow}, \\ 
\hat{\cal V}&=&V\sum_{i,\sigma,\sigma^{\prime}}n_{i,\sigma}n_{i+1,\sigma^{\prime}}, 
\end{eqnarray}
where $C^{\dagger}_{i,\sigma}$ is the creation 
operator of $\sigma$ spin electron at $i$ site, 
$n_{i,\sigma}$ is the number operator, $g=2$, 
$i=1,\cdots,N_{\rm S}$, $N_{\rm S}$ is the 
number of the total sites and $\sigma =\uparrow$ and $\downarrow$. 
When the magnetic field is 
applied to (x or z)-axis ($H_x$ or $H_z$), 
$H_{j}=H\hat{\sigma}_{j}$ ($j=x$ and $z$), where 
$H$ is the strength of the magnetic field and 
$\hat{\sigma}_{j}$ is Pauli spin matrix. 
In this model, the filling of electrons is 1/4. 


The interaction term, $\hat{\cal U}$ and 
$\hat{\cal V}$ are treated in mean field
approximation as

\begin{eqnarray}
\hat{\cal U}_{\rm M}&=&\sum_{k_{x}}
\sum_{Q}\{
\rho_{\uparrow}(Q)
C^{\dagger}(k_{x},
\downarrow) 
C(k_{x}-Q,\downarrow)  \nonumber \\
&+&\rho^{*}_{\downarrow}(Q)
C^{\dagger}(k_{x}-Q,
\uparrow) 
C(k_{x},\uparrow)\} \nonumber \\
&-&\frac{1}{I}\sum_{Q}\rho_{\uparrow}(Q)\rho^{*}_{\downarrow}(Q), \\
\hat{\cal V}_{\rm M}&=&(\frac{V}{U})\sum_{k_{x},
\sigma,\sigma^{\prime}}\sum_{Q}e^{-iQa}\{
\rho_{\sigma}(Q)
C^{\dagger}(k_{x}, 
\sigma^{\prime}) 
C(k_{x}-Q,
\sigma^{\prime})  \nonumber \\
&+&\rho_{\sigma^{\prime}}^{*}(Q)
C^{\dagger}(k_{x}, 
\sigma) 
C(k_{x}-Q,
\sigma)\} \nonumber \\
&-&\frac{V}{IU}\sum_{Q,\sigma,\sigma^{\prime}}e^{-iQa}\rho_{\sigma}(Q)
\rho_{\sigma^{\prime}}^{*}(Q),
\end{eqnarray}
where $I=U/N_{\rm S}$. 
The self-consistent equation for the order parameter $\rho_{\sigma}(Q)$ 
is given by 
\begin{eqnarray}
\rho_{\sigma}(Q)&=&I\sum_{k_{x}}
<C^{\dagger}(k_{x},
\sigma)
C(k_{x}-Q,
\sigma)>.
\end{eqnarray}
We use the mean field, $\rho_{\sigma}(Q)$, by 
the coupling between electrons with same spins. 
In order to simplify, we do not consider the case of 
the mean field, 
$\bar{\rho_{\sigma}}(Q)=I\sum_{k_{x}}<C^{\dagger}(k_{x},\sigma)C(k_{x}-Q,\bar{\sigma})>$, by the coupling of electrons with opposite spin.

We limit the sum of the wave vector as $Q=q,2q,3q$ and $4q$ 
($q=2k_{\rm F}$), because 
the wave vectors of $2k_{\rm F}=\pi /2a$ and its higher harmonics 
should be considered due to the nesting of the Fermi surface 
in the one-dimensional quarter filled band. 
We can obtain the self-consistent solutions from eq. (8) by 
using eigenvectors obtained by diagonalizing 
$\hat{\cal K}+\hat{\cal U}_{\rm M}+\hat{\cal V}_{\rm M}$, 
which becomes $8\times 8$ matrix. 
The electron density at $j$ site, $n(j)$, and the 
spin moment at $j$ site, $S_z(j)$, are given by 
\begin{eqnarray}
n(j)&=&\frac{1}{U}\sum_{Q,\sigma}\rho_{\sigma}(Q)e^{iQja}, \\
S_z(j)&=&\frac{1}{2U}\sum_{Q}(\rho_{\uparrow}(Q)-\rho_{\downarrow}(Q))e^{iQja}.
\end{eqnarray}
The notation in this paper follows Seo and Fukuyama.\cite{seofukuyama} 
We can calculate 
the total energy, $E$,
\begin{eqnarray}
E=\sum_{i=1,\sigma}\epsilon_{i,\sigma},
\end{eqnarray}
where the sum is limited to electron filling and 
$\epsilon_{i,\sigma}$ is an eigenvalue. 
From the ordered state energy ($E_{\rm OS}$) and 
the normal state energy ($E_{\rm N}$), 
the energy gain ($E_{\rm g}$) can be obtained 
by $E_{\rm g}=E_{\rm OS}-E_{\rm N}$.

The Pauli spin susceptibility, $\chi$, is 
$\mu_{\rm B}^2N(0)$ ($N(0)$ is the density of state 
on the Fermi energy and 
$N(0)=N_{\rm S}/(4\pi t\sin ak_{\rm F})$),\cite{pauli,pauli2,Mckenzie} 
and the energy gain from 
the normal state is given by 
$-\chi H^2$. 
When $E_{\rm g}=-\chi H_p^2$, 
the Pauli limit field, $h_p$, is given by 
\begin{eqnarray}
h_p=2\sqrt{\frac{-\pi E_{\rm g}}{\sqrt{2}N_{\rm S}}},
\end{eqnarray}
where $h_p=\mu_{\rm B}gH_p/2t$.



\begin{figure}
\leavevmode
\epsfxsize=8cm
\epsfbox{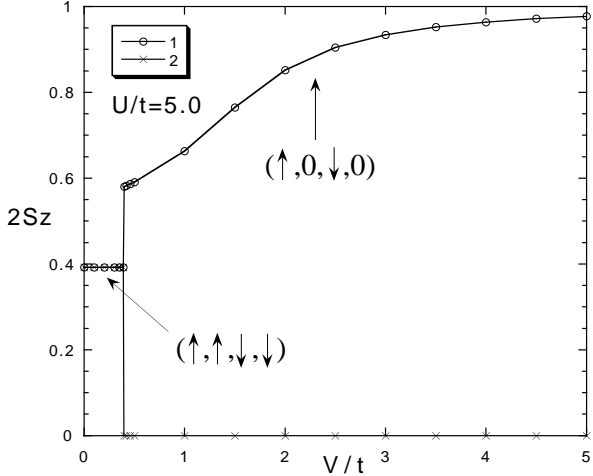}
\caption{2S$_z$ as a function of $v$ at $H=0$ 
}
\label{fig:1}
\end{figure}

\begin{figure}
\leavevmode
\epsfxsize=7cm
\epsfbox{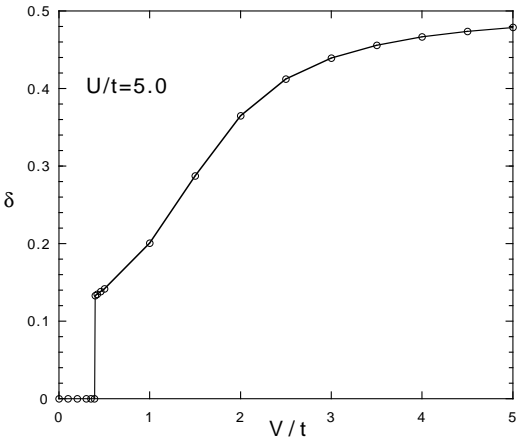}
\caption{$\delta$ as a function of $v$ at $H=0$ 
}
\end{figure}

\section{Results and Discussions}
\subsection{Strong coupling}
First, we show the result at $H=0$ 
at $U/t=5.0$, whose large value 
means the strongly correlated system. 
Figs. 1 and 2 are $S_z$ and $\delta$ 
as a function of $V/t$ 
at $U/t=5.0$. 
At $0\leq V\leq 0.392$, 
the antiferromagnetic ordering 
(($\uparrow$,$\uparrow$,$\downarrow$,$\downarrow$), i.e., 
$S_z(1)=S_z(2)=-S_z(3)=-S_z(4)$) 
is stabilized and there is no charge ordering. 
The spin ordering of 
($\uparrow$,$\uparrow$,$\downarrow$,$\downarrow$) has 
the wave vector of $2k_{\rm F}$. 
Above $V/t=0.392$, 
the spin ordering becomes 
($\uparrow$,0,$\downarrow$,0) 
($S_z(1)=-S_z(3)$, $S_z(2)=S_z(4)=0$) 
and the charge ordering ($\delta$,-$\delta$,$\delta$,-$\delta$) exist, 
where 
$n(1)=n(3)$=0.5+$\delta$, $n(2)=n(4)$=0.5-$\delta$, 
which can be seen in Figs. 1 and 2. 
These ($\uparrow$,0,$\downarrow$,0) and ($\delta$,$-\delta$,$\delta$,$-\delta$) 
mean $2k_{\rm F}$-SDW and $4k_{\rm F}$-CDW, respectively. 
By including the inter-site Coulomb interaction, 
$4k_{\rm F}$-CDW is induced.

When $U/t$ and $V/t$ are smaller than ($\sim 1.5$), 
we understand that the system with small $S_z$ is 
the SDW transition due to Peierls instability 
of the Fermi surface. 
The spin and charge orderings are localized 
if $U/t$ and $V/t$ become larger than ($\sim 4.0$) since 
the amplitudes of $S_z$ and $n$ are saturated. 
This state with the localized spin 
and charge orderings is 
Mott antiferromagnetic state due to the 
larger values of $U/t$ and $V/t$. 
These are the same results as Seo and Fukuyama.\cite{seofukuyama}

Next, we show 
$S_z$ and $n$ at $H\neq0$ and $U/t=V/t=5.0$ 
by using the ground state, ($\uparrow$,0,$\downarrow$,0) 
and ($\delta$,-$\delta$,$\delta$,-$\delta$), 
in the strong coupling system. 
When the magnetic field is applied along 
$x$-axis ($h_x=\mu_{\rm B}gH_{x}/2t$), 
the antiferromagnetic state is gradually suppressed up to 
the critical field ($h_x^{c}=2.4$) and above $h_x^{c}$ 
the spin ordering becomes $(0, 0, 0, 0)$, 
whereas the charge ordering is unchanged, 
as shown in Figs. 3 and 4, where 
$S_z(1)=-S_z(3)$, $S_z(2)=S_z(4)$ and 
$n(1)=n(3)$=0.5+$\delta$, $n(2)=n(4)$=0.5-$\delta$. 
The $h_x$-dependence of the amplitude of 
$S_z$ is in good agreement with eq. (1) when we set 
$H_x^0$ as $H_x^c=2th_x^c/\mu_{\rm B}g$, 
which is shown by solid lines 
in Fig. 3.

In the case of $h_z=\mu_{\rm B}gH_{z}/2t\neq 0$, 
the alignment of the spin moment, ($\uparrow$,0,$\downarrow$,0) when 
$h_z^{c}=0$ is kept as $h_z^{c}$ increases, but, 
the system becomes 
($\downarrow$,0,$\downarrow$,0) 
at $h_z^{c}=2.4$, 
as shown in Fig.5. 
This $h_z^{c}$ has the same value of $h_x^{c}$. 
The charge ordering is not changed upon 
increasing $h_z$, as shown in Fig. 6, where 
$n(1)=n(3)$=0.5+$\delta$, $n(2)=n(4)$=0.5-$\delta$. 
These ($\downarrow$,0,$\downarrow$,0) and 
($\delta$,$-\delta$,$\delta$,$-\delta$) are $4k_{\rm F}$-SDW 
and $4k_{\rm F}$-CDW, that is, 
above $h_z^{c}$ the system becomes 
paramagnetic state to stay to be 
localized due to larger $U/t$ and $V/t$.

The $h_x$- and $h_z$-
dependences of the spin ordering in 
our results can be understood 
by the mean field solutions for 
strong coupled Ising 
model mentioned 
in the introduction.\cite{ising}

The charge ordering is unchanged by the magnetic 
field. The magnetic field is not contributed 
to localized electrons made by 
larger $V/t$.







\begin{figure}
\leavevmode
\epsfxsize=7cm
\epsfbox{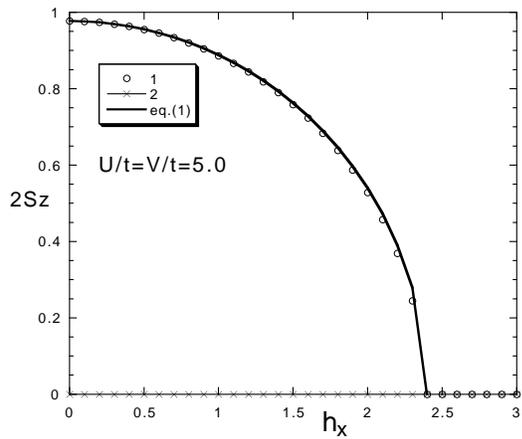}
\caption{
When $U/t=V/t=5.0$, 
2S$_z$ as a function of $h_x$. 
The solid lines are written by 
eq. (1). 
}
\end{figure}

\begin{figure}
\leavevmode
\epsfxsize=7cm
\epsfbox{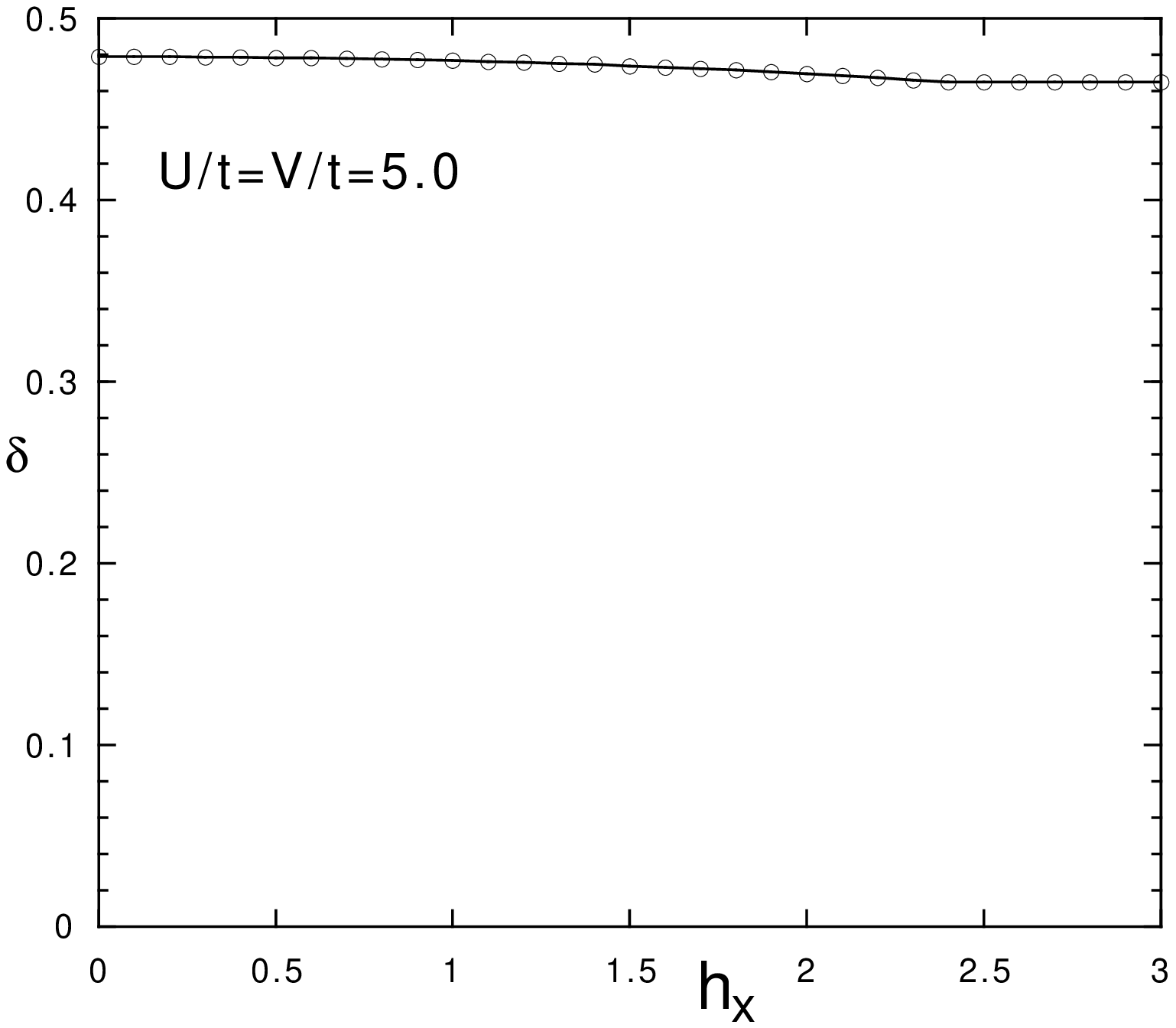}
\caption{
When $U/t=V/t=5.0$, 
$\delta$ as a function of $h_x$. 
}
\end{figure}

\begin{figure}
\leavevmode
\epsfxsize=7cm
\epsfbox{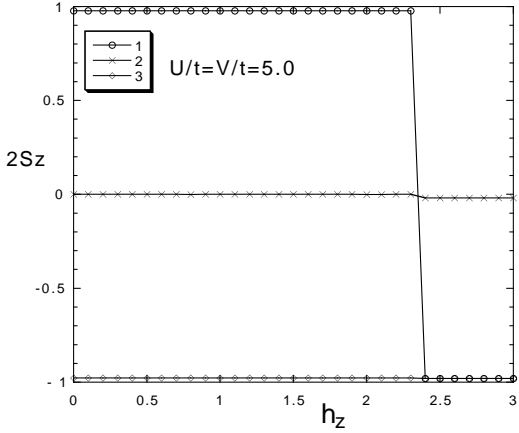}
\caption{
When $U/t=V/t=5.0$, 
2S$_z$ as a function of $h_z$. 
}
\end{figure}

\begin{figure}
\leavevmode
\epsfxsize=7cm
\epsfbox{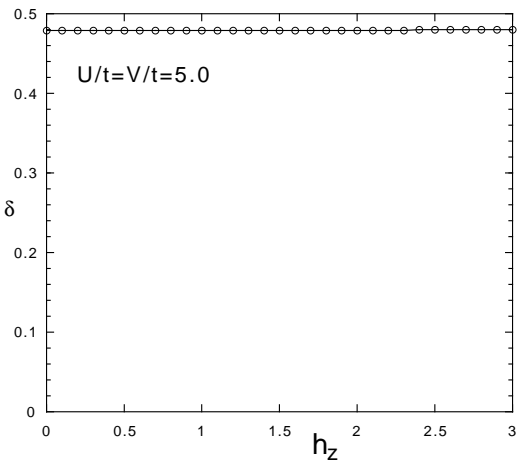}
\caption{
When $U/t=V/t=5.0$, 
$\delta$ as a function of $h_z$. 
}
\end{figure}

\subsection{Non-Strong coupling}
We calculated the solutions as a function of $V/t$ at 
$U/t=1.5$. There were two solutions 
(($\uparrow$,$\uparrow$,$\downarrow$,$\downarrow$) and 
($\uparrow$,0,$\downarrow$,0) and 
($\delta$,$-\delta$,$\delta$,$-\delta$)), 
which are shown 
in Figs 7, 8 and 9. 
These are for the state with 
$2k_{\rm F}$-SDW and the coexistent state with 
$2k_{\rm F}$-SDW and $4k_{\rm F}$-CDW, respectively. 
It is found that these amplitudes of $S_z$ and 
$\delta$ are small. 
In the region of 
$0\leq V/t\leq 1.5$, 
the energies with these solutions are nearly 
the same, namely, these states are degenerate. 
Therefore, we analyze 
$h_x$- and $h_z$-dependences of $S_z$ and $n$ 
for two solutions.

We calculate the case of $U/t=V/t=1.5$ 
at $H \neq0$. 
On using 
($\uparrow$,$\uparrow$,$\downarrow$,$\downarrow$) when $H=0$, 
for $h_{z}$ 
this state is changed to 
($\downarrow$,$\downarrow$,$\downarrow$,$\downarrow$) at 
$h_{z}^{c}=0.0155$, but, 
for $h_{x}$, $S_z$ is unchanged, which are shown 
in Figs. 10 and 11.
This suppression for $h_{z}$ comes from 
Pauli paramagnetic limit since 
$h_{z}^{c}$ corresponds to $h_p\simeq 0.0125$ 
obtained by using eq. (12) and $E_{\rm g}/N_{\rm S}\simeq0.0000175$ 
calculated at 
$U/t=V/t=1.5$ and $H=0$. 
For $h_{x}$, there is no Pauli limit, 
because the nesting vector of the SDW is not 
affected by Zeeman splitting. 
In Fig. 11, $S_z$ linearly increases as $h_z$ 
increases, which means that 
the paramagnetic state is stabilized by the magnetic 
field.

For ($\uparrow$,0,$\downarrow$,0) and 
($\delta$,-$\delta$,$\delta$,-$\delta$), 
when the magnetic field is applied along $x$-axis, 
the spin and charge ordering do not change, as shown in 
Figs. 12 and 13. 
However, for $h_{z}$, both orderings of 
the spin and charge disappear at $h_{z}^{c}=0.0155$, 
which is corresponding to Pauli paramagnetic limit field, 
$h_p\simeq 0.0125$, as shown in 
Figs. 14 and 15. 
This is due to the effect of 
Pauli limit, too. 
The coexistent state of 
$2k_{\rm F}$-SDW and $4k_{\rm F}$-CDW 
with ($\uparrow$,0,$\downarrow$,0) 
and 
($\delta$,-$\delta$,$\delta$,-$\delta$) is 
changed to paramagnetic state 
with ($\downarrow$,$\downarrow$,$\downarrow$,$\downarrow$) 
and (0,0,0,0). 
It is seen that the linear increasing of $S_z$ 
above $h_{z}^{c}$.


\begin{figure}
\leavevmode
\epsfxsize=7cm
\epsfbox{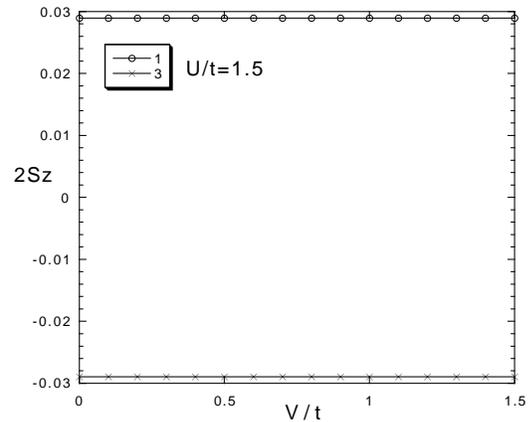}
\caption{By using of 
($\uparrow$,$\uparrow$,$\downarrow$,$\downarrow$) 
at $U/t=1.5$ and $V/t=0$, 
$2S_z$ as a function of $v$ at $H=0$. 
}
\end{figure}

\begin{figure}
\leavevmode
\epsfxsize=7cm
\epsfbox{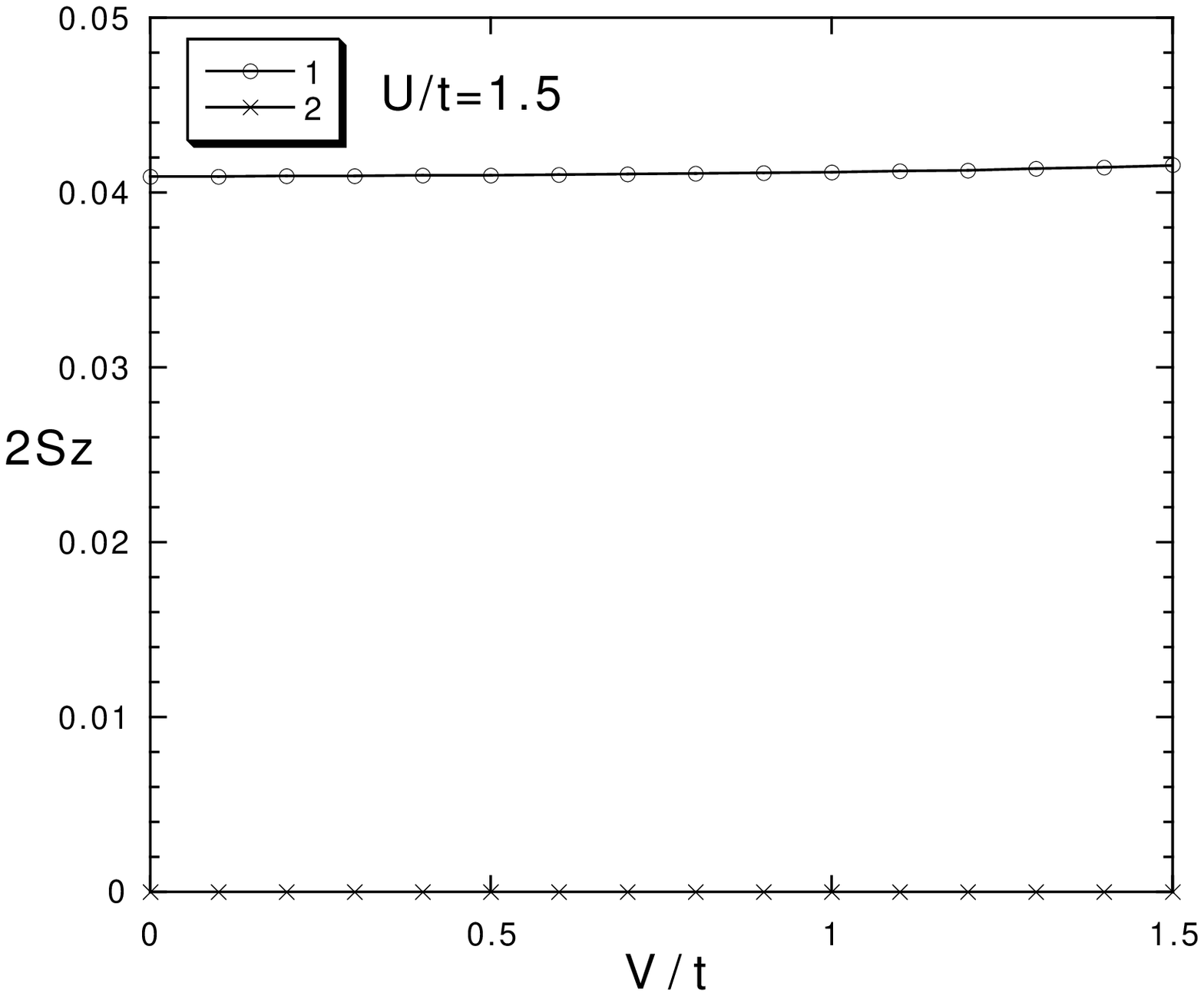}
\caption{By using of 
($\uparrow$,0,$\downarrow$,0) 
at $U/t=1.5$ and $V/t=0$, 
$2S_z$ as a function of $v$ at $H=0$. 
}
\end{figure}

\begin{figure}
\leavevmode
\epsfxsize=7cm
\epsfbox{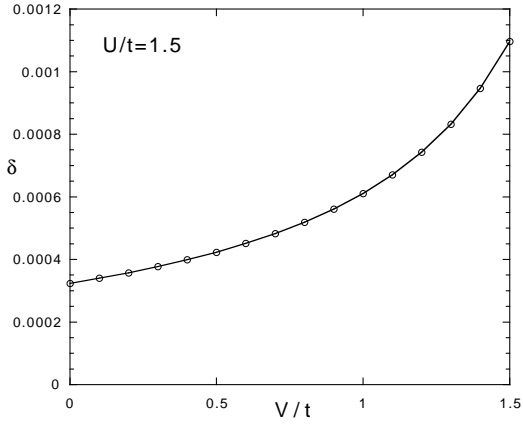}
\caption{By using of 
($\uparrow$,0,$\downarrow$,0) 
at $U/t=1.5$ and $V/t=0$, 
$\delta$ as a function of $v$ at $H=0$. 
}
\end{figure}

\begin{figure}
\leavevmode
\epsfxsize=7cm
\epsfbox{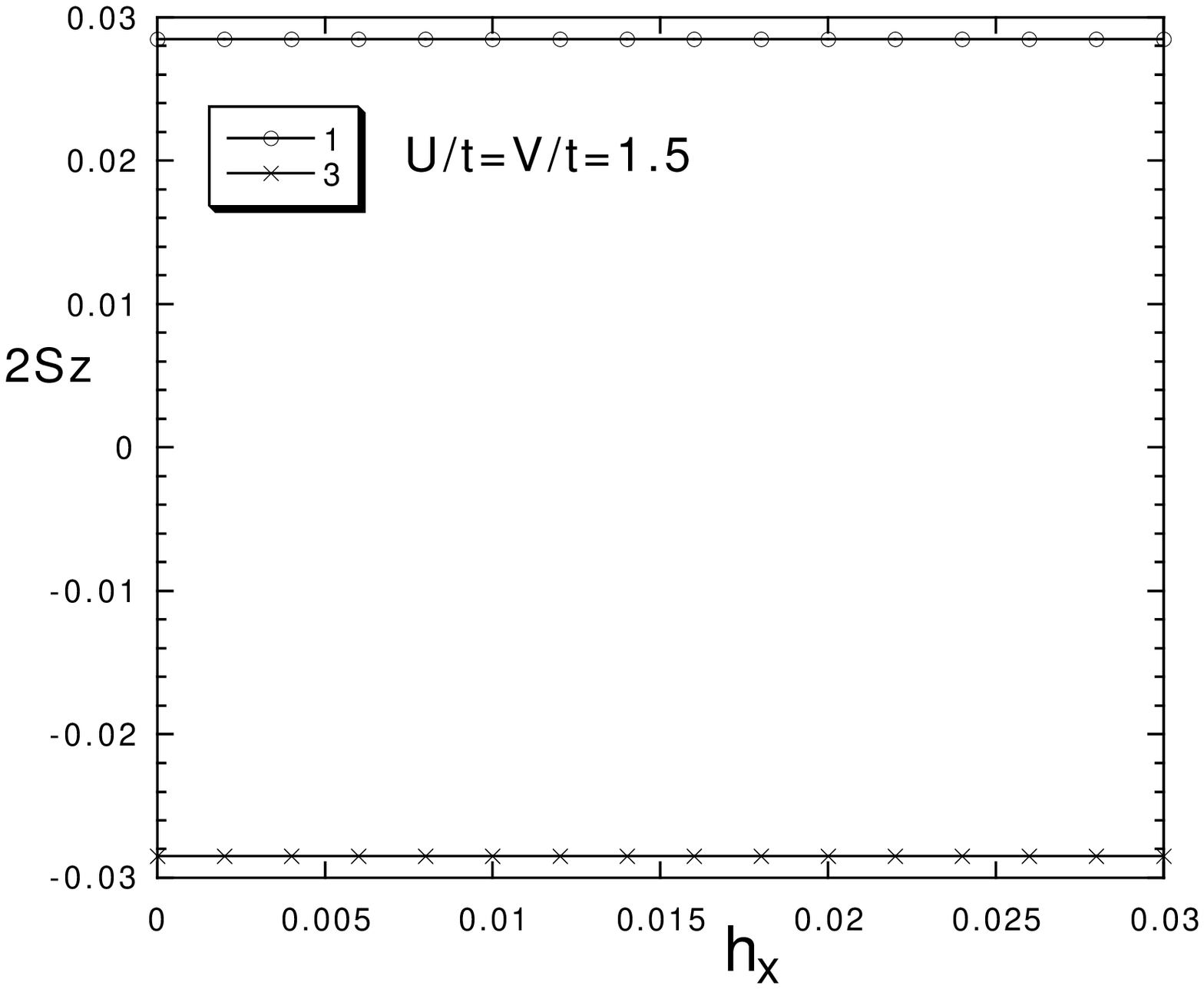}
\caption{
When $U/t=V/t=1.5$, 
$2S_z$ as a function of $h_x$ by using of 
($\uparrow$,$\uparrow$,$\downarrow$,$\downarrow$) 
at $h_x=0$.
}
\end{figure}

\begin{figure}
\leavevmode
\epsfxsize=7cm
\epsfbox{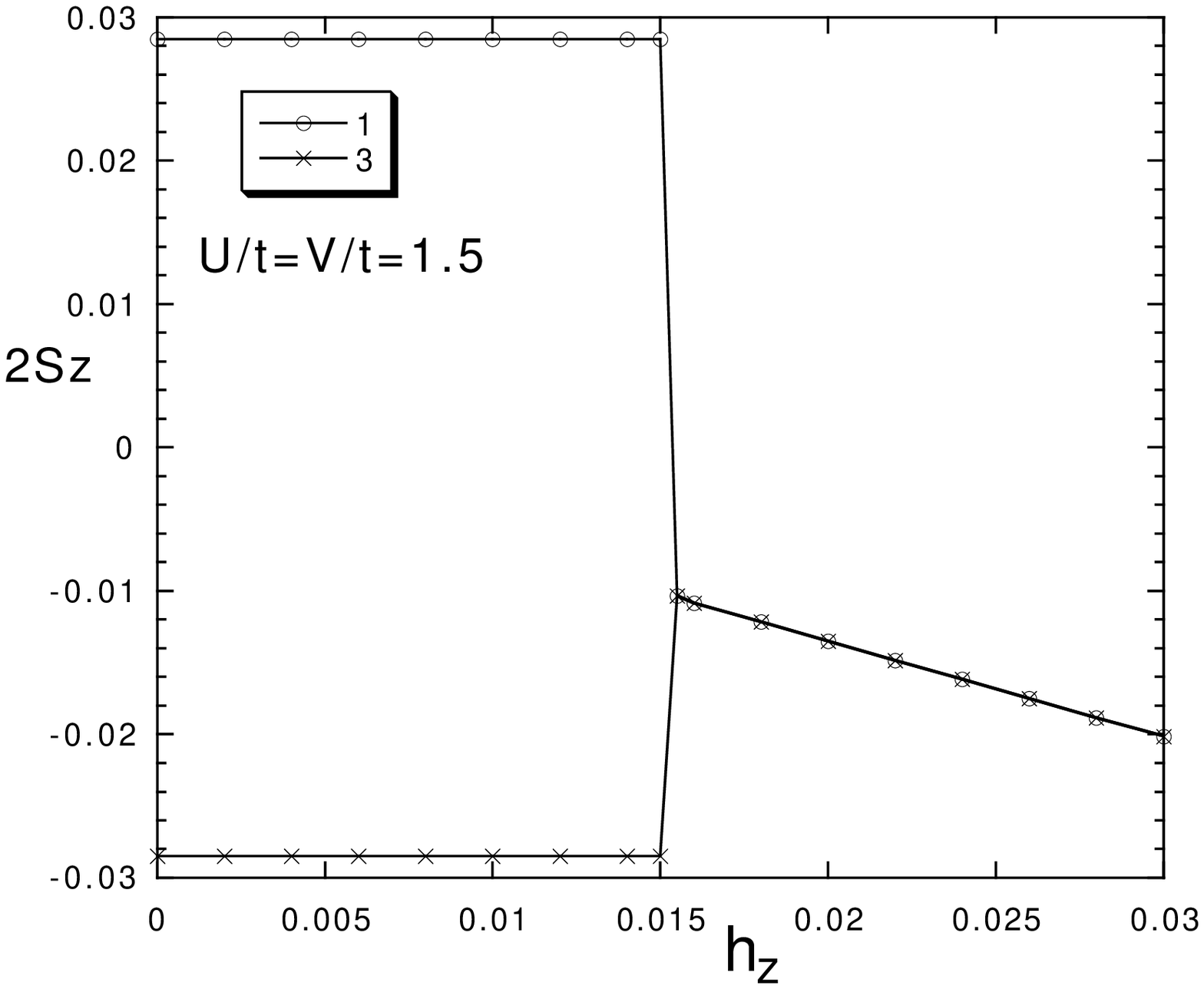}
\caption{
When $U/t=V/t=1.5$, 
$2S_z$ as a function of $h_z$ by using of 
($\uparrow$,$\uparrow$,$\downarrow$,$\downarrow$) 
at $h_z=0$.
}
\end{figure}

\begin{figure}
\leavevmode
\epsfxsize=7cm
\epsfbox{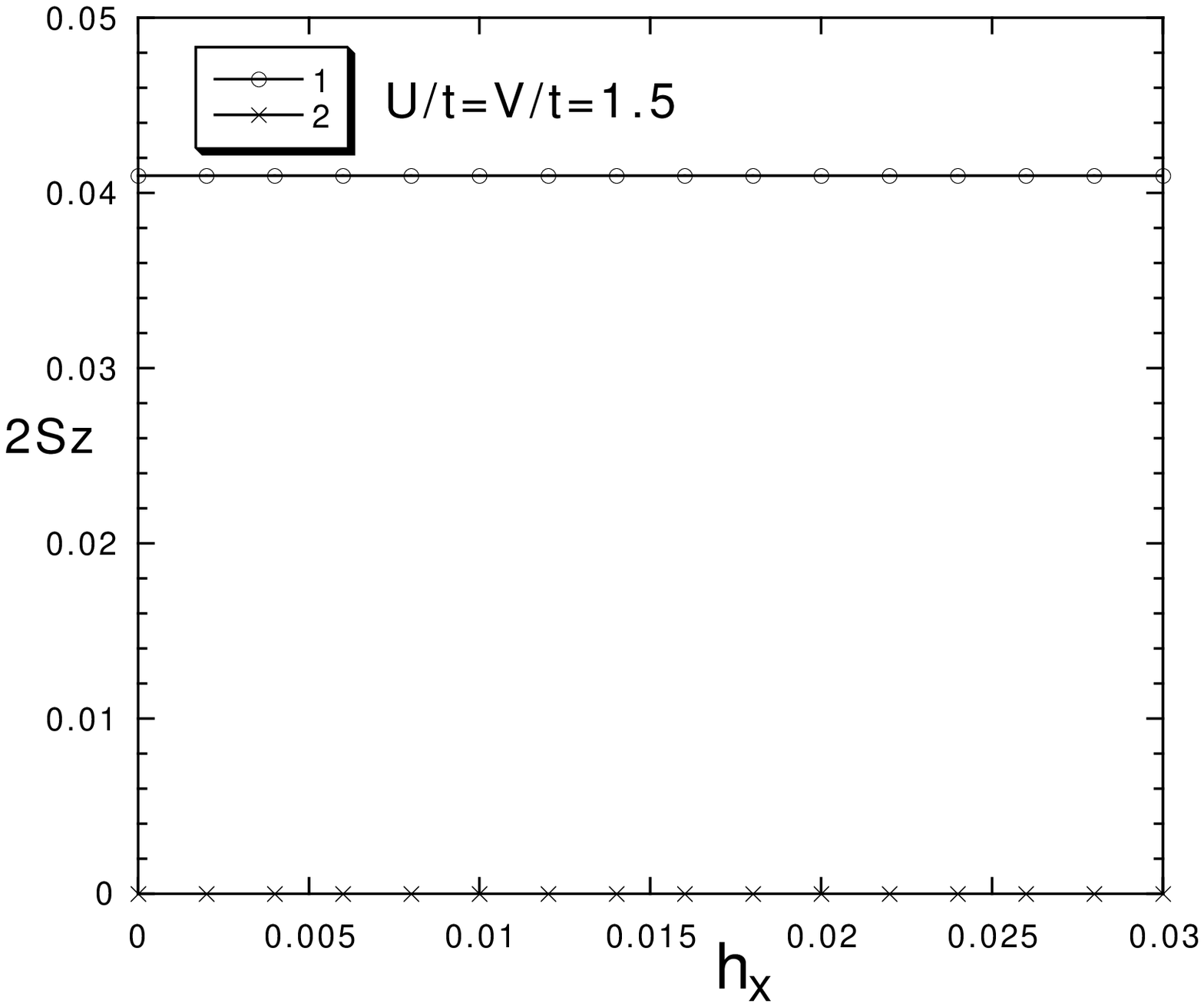}
\caption{
When $U/t=V/t=1.5$, 
$2S_z$ as a function of $h_x$, 
by using of 
($\uparrow$,0,$\downarrow$,0) 
at $h_x=0$.
}
\end{figure}

\begin{figure}
\leavevmode
\epsfxsize=7cm
\epsfbox{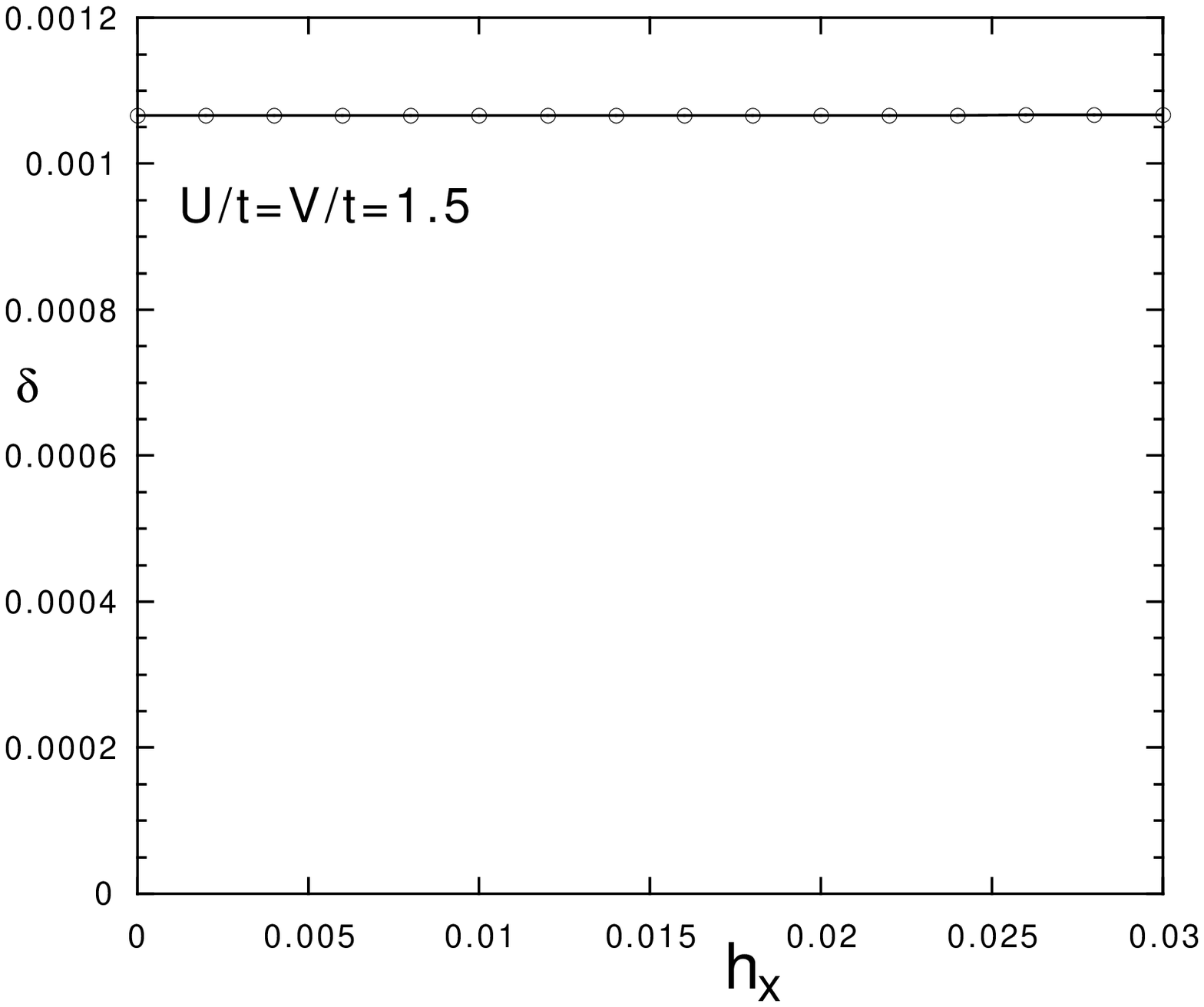}
\caption{
When $U/t=V/t=1.5$, 
$\delta$ as a function of $h_x$, 
by using of 
($\uparrow$,0,$\downarrow$,0) 
at $h_x=0$.
}
\end{figure}

\begin{figure}
\leavevmode
\epsfxsize=7cm
\epsfbox{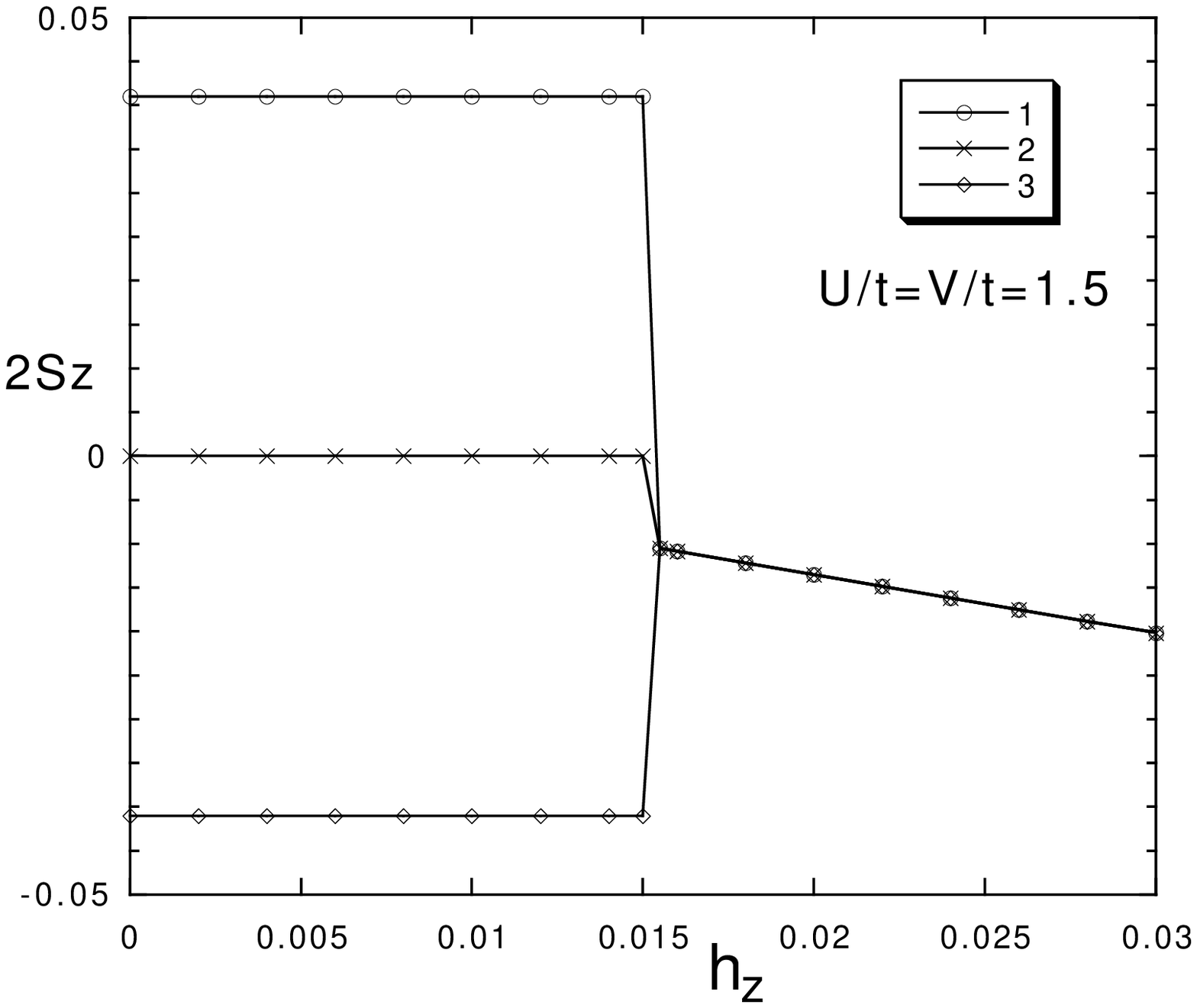}
\caption{
When $U/t=V/t=1.5$, 
$2S_z$ as a function of $h_z$, 
by using of 
($\uparrow$,0,$\downarrow$,0) 
at $h_z=0$.
}
\end{figure}

\begin{figure}
\leavevmode
\epsfxsize=7cm
\epsfbox{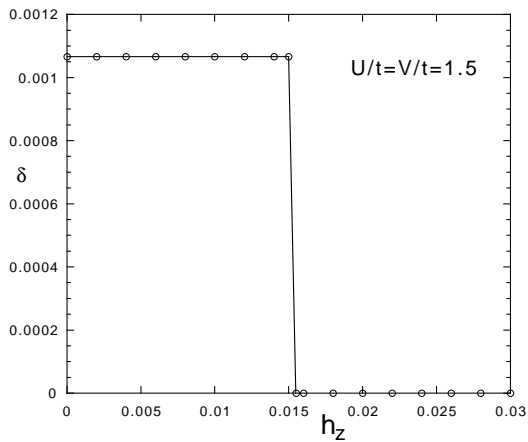}
\caption{
When $U/t=V/t=1.5$, 
$\delta$ as a function of $h_z$, 
by using of 
($\uparrow$,0,$\downarrow$,0) 
at $h_z=0$.
}
\end{figure}


\subsection{Comparison with Experiments}
The coexistent state with 
CDW and SDW is realized due to the 
inter-site Coulomb interaction even if 
the correlation between electrons 
is strong or not.

In the non-localized SDW system, 
when the magnetic field is applied parallel to 
the easy axis, both orderings of SDW and CDW disappear at 
the critical field of the Pauli paramagnetic limit. 
In the case of the magnetic field perpendicular to 
easy axis, both orderings of 
CDW and SDW are unchanged. 
In (TMTSF)$_2$$X$, for example, since 
the easy axis is $b$-axis, it is expected 
that when the magnetic 
field is applied to $b$-axis 
both orderings become to disorder at 
Pauli limit field.

On the other hand, 
in the localized antiferromagnetic state such 
as (TMTTF)$_2$$X$, the charge ordering is not 
suppressed in both cases of 
the magnetic field applied to parallel to and 
perpendicular to the easy axis. 
Thus, 
$4k_{\rm F}$-CDW may be 
observed from the X-ray measurement even if 
the magnetic field is applied to $a$, $b$ and $c$-axis. 
It is a means of finding whether the system 
is strongly correlated or not. 

In (DCNQI)$_2$Ag, which are strongly correlated system such 
as (TMTTF)$_2$$X$, $4k_{\rm F}$-CDW has been observed 
at zero magnetic field.\cite{hiraki,moret} 
Even under high fields, the charge ordering 
should be appeared.

\section{Conclusions}
We theoretically study the coexistent state 
of CDW and SDW under the magnetic field. 
As a result, in the case of the strongly coupling system, 
although the spin ordering is suppressed at 
high fields, the charge ordering still remains. 
When the coupling is not so large, 
the CDW and the SDW disappear at Pauli paramagnetic limit field. 

These features of the coexisitent state of 
CDW and SDW under 
magnetic fields should be observed in 
the strongly correlated system (non-strongly correlated system) 
such as (TMTTF)$_2$$X$ and (DCNQI)$_2$Ag ((TMTSF)$_2$$X$).



\section{Acknowledgment}

One of the authors (K. K.) would like to thank 
T. Sakai for valuable discussions.
K. K. was partially supported by Grant-in-Aid for
JSPS Fellows from the Ministry of
Education, Science, Sports and Culture.
K. K.
was financially supported by the Research Fellowships
of the Japan Society
for the Promotion of Science for Young Scientists.

\end{document}